\documentclass[aps,prb,reprint,groupedaddress,showpacs,superscriptaddress,
]{revtex4-1}
\usepackage{hyperref}
\usepackage{amssymb}
\usepackage{graphicx}
\usepackage{dcolumn}
\usepackage{bm}
\usepackage{amsmath}
\usepackage{fixmath}
\usepackage[usenames]{color}
\usepackage{float}
\usepackage{bbm}
\usepackage{epsfig}
\newcommand{\td}{t_{\text{diag}}}
\newcommand{\beq}{\begin{equation}}
\newcommand{\eeq}{\end{equation}}
\newcommand{\beqa}{\begin{eqnarray}}
\newcommand{\eeqa}{\end{eqnarray}}
\newcommand{\ba}{\begin{align}}
\newcommand{\ea}{\end{align}}
\newcommand{\etal}{\emph{et al.}}

\newcommand{\iwn}{\text{i}\omega_n}
\newcommand{\BK}{{\bm{k}}}

\preprint{\fbox{\texttt{\tiny\jobname.tex}}}
\begin{document}
\title{Charge and spin criticality for the continuous Mott transition in a two-dimensional organic conductor
}
\author{Michael Sentef}
\email[]{sentefmi@physik.uni-augsburg.de}
\affiliation{Theoretical Physics III, Center for Electronic Correlations and 
Magnetism, Institute of Physics, University of Augsburg, D-86135 Augsburg,
Germany}
\affiliation{Stanford Institute for Materials and Energy Science,
SLAC National Accelerator Laboratory, 2575 Sand Hill Road, Menlo Park, CA 94025, USA}
\author{Philipp Werner}
\affiliation{Theoretische Physik, ETH Zurich, 8093 Z{\"u}rich, Switzerland}
\author{Emanuel Gull}
\affiliation{
Department of Physics, Columbia University, New York, NY 10027, USA}
\author{Arno P.\ Kampf}
\affiliation{Theoretical Physics III, Center for Electronic Correlations and 
Magnetism, Institute of Physics, University of Augsburg, D-86135 Augsburg,
Germany}
\date{\today}
\begin{abstract}
We study the continuous bandwidth-controlled Mott transition in the 
two-dimensional single-band Hubbard model with a focus on the critical scaling 
behavior of charge and spin degrees of freedom. Using plaquette cluster dynamical mean-field theory, we find charge and spin criticality consistent with experimental 
results for organic conductors. In particular, the charge degree of freedom 
measured via the local density of states at the Fermi level shows a smoother 
transition than expected for the Ising universality class and in single-site 
dynamical mean-field theory, revealing the importance of short-ranged nonlocal correlations in two spatial dimensions. The spin criticality measured via the local spin 
susceptibility agrees quantitatively with nuclear magnetic resonance 
measurements of the spin-lattice relaxation rate. 
\end{abstract}
\maketitle
The Mott metal-insulator transition (MIT) is a paradigmatic example of a 
correlation-induced phase transition.\cite{Imada98} Its physics is generically
contained in the single-band Hubbard model, which is parametrized by the local 
Coulomb repulsion $U$, the bare bandwidth $W$, and the average electron density
$n$. Two MITs are distinguished: first, the bandwidth-controlled Mott 
transition at fixed filling, where an insulator turns into a metal by 
increasing $W/U$, typically realized in experiments through chemical or 
hydrostatic pressure, and second the filling-controlled Mott transition at 
fixed $U/W$, where the system becomes metallic upon adding electrons or holes. 
A paramagnetic Mott transition is often superceded by antiferromagnetic 
ordering unless the system is frustrated or the temperature high enough so that 
the bandwidth-controlled MIT proceeds from a paramagnetic metal to a 
paramagnetic insulator. Typically, the paramagnetic MIT is a discontinuous 
first-order transition at low temperatures with a continuous critical end point.

The scaling behavior at the critical end point, or in short Mott criticality, 
for the bandwidth-controlled MIT has been probed experimentally for the charge 
degree of freedom via the dc conductivity, e.g. for (V$_{1-x}$Cr$_x$)$_2$O$_3$ 
\cite{Limelette03} or the quasi-two-dimensional 
$\kappa$-(ET)$_2$Cu[N(CN)$_2$]Cl (abbreviated as $\kappa$-Cl).\cite{Limelette03org,Kagawa05} The latter belongs to a class of layered 
organic charge transfer salts,\cite{Kino96,Dumm09} which are both 
low-dimensional and geometrically frustrated, such that magnetic order is 
suppressed at the temperature of the critical end point of the Mott transition. These organic salts therefore allow to follow the first-order MIT to its second order critical end point in the absence of magnetic long-range order. Only recently Mott 
criticality was also investigated for the spin degree of freedom by nuclear 
magnetic resonance (NMR) measurements under pressure.\cite{Kagawa09} The focus
of the latter study was on the critical scaling behavior upon varying pressure 
at fixed temperature, which is described by a critical exponent $\delta$. Here 
we present a theoretical modeling of the spin and charge Mott criticality and 
determine $\delta$. The spin criticality will be investigated via the local 
spin susceptibility, which is related directly to the NMR spin-lattice 
relaxation rate $1/T_1$. For the charge criticality we focus on the local density of 
states at the Fermi energy. 

Experimentally Mott criticality is probed by the scaling behavior of a selected
quantity $\sigma$ (e.g., the conductivity) as a function of external parameters
such as the temperature $T$ or the  pressure $p$ near the critical end point. 
Specifically one observes scaling with respect to the reduced parameters 
$t_{\text{red}}$ $=$ $(T-T_c)/T_c$ and $p_{\text{red}}$ $=$ $(p-p_c)/p_c$, where 
the index $c$ denotes the values at the critical end point. Criticality is then
classified by the set of exponents $\beta$, $\gamma$, and $\delta$ via
\beqa
\sigma (t_{\text{red}}, p_{\text{red}}=0) - \sigma_c &\propto& 
|t_{\text{red}}|^{\beta},\nonumber \\ 
\frac{\partial \sigma(t_{\text{red}},p_{\text{red}})}{\partial 
p_{\text{red}}}\Big|_{p_{\text{red}}=0} &\propto& |t_{\text{red}}|^{-\gamma},
\nonumber \\ 
\sigma (t_{\text{red}}=0, p_{\text{red}}) - \sigma_c  &\propto& 
|p_{\text{red}}|^{1/\delta},
\label{eq:exponents}
\eeqa
where $\sigma_c$ $=$ $\sigma(t_{\text{red}}=0,p_{\text{red}}=0)$. The critical 
exponents obey the scaling law $\gamma$ $=$ $\delta$ 
($\beta-1$).\cite{Goldenfeld}

The Mott transition has been proposed to be in the Ising universality 
class \cite{Castellani79,Kotliar00,Onoda03,Papanikolaou08} based on the 
assumption that the double occupancy may serve as a fingerprint observable for 
the MIT, which plays a  similar role as the order parameter in a thermodynamic phase transition to a broken symmetry state. The universality class of a second-order phase 
transition is determined by the symmetry of the order parameter and the spatial
dimension; the scalar character of the double occupancy would imply the Ising 
universality class.\cite{Goldenfeld} Conductivity measurements for 
(V$_{1-x}$Cr$_x$)$_2$O$_3$ indeed confirm critical behavior compatible with 3D 
Ising universality ($\beta$ $\approx$ $0.33$, $\gamma$ $\approx$ $1.2$, 
$\delta$ $\approx$ $4.8$),\cite{Limelette03,Goldenfeld} but the situation for 
the two-dimensional $\kappa$-Cl has remained controversial.

The critical exponents of the 2D Ising model are $\beta$ $=$ $1/8$, $\gamma$ 
$=$ $7/4$, and $\delta$ $=$ $15$.\cite{McCoy73} Conductivity measurements 
under pressure performed on $\kappa$-Cl challenge the prediction of Ising 
universality, since the observed exponents are $\beta$ $\approx$ $1$, $\gamma$ 
$\approx$ $1$, and $\delta$ $\approx$ $2$. Imada {\etal} argued that the 
observed deviation from Ising universality is a manifestation of unconventional
quantum criticality specific to a two-dimensional system.\cite{Imada05,Imada10} A different scenario was proposed by Papanikolaou 
{\etal} who claimed that the conductivity can have a different critical 
behavior than the order parameter of the transition.\cite{Papanikolaou08}

In NMR experiments under pressure on $\kappa$-Cl Kagawa {\etal} observed that 
the critical enhancement of the conductivity upon passing through the critical 
end point is accompanied by a critical suppression of spin fluctuations;\cite{Kagawa09} the latter was inferred from a decrease of the nuclear 
spin-lattice relaxation rate $T_1^{-1}$. Identical critical exponents $\delta$ 
were determined for the conductivity and the spin relaxation $1/(T_1T)$ within 
experimental accuracy. 

Here we aim at a microscopic description of Mott criticality in organic 
conductors by studying the two-dimensional one-band Hubbard model on an 
anisotropic triangular lattice with the Hamiltonian
\begin{equation}
H = \sum_{\BK,\sigma} (\epsilon_{\BK}-\mu)
c^\dagger_{\BK,\sigma}c^{}_{\BK,\sigma}+U\sum_i n_{i,\uparrow}n_{i,\downarrow},
\end{equation} 
where $c^\dagger_{\BK,\sigma}$ ($c^{}_{\BK,\sigma}$) creates (annihilates) an 
electron in a Bloch state with lattice momentum $\BK$. $n_{i,\sigma}$ is the 
local density operator for site $i$ and spin $\sigma$ $=$ 
$\uparrow,\downarrow$, $U$ $>$ 0 is the local Coulomb repulsion strength, and 
$\mu$ is the chemical potential. The electronic dispersion is given by
\begin{equation}
\epsilon_\BK=-2t\left(\cos k_x + \cos k_y \right)-2\td \cos(k_x+k_y).
\end{equation}
Following Ref.\ \onlinecite{Kandpal09} we choose for $\kappa$-Cl a diagonal 
hopping $\td=0.44 t$ and fix the filling at $n=1$ in a grand-canonical 
calculation. 

We obtain an approximate solution of the Hubbard model by using cluster dynamical mean-field theory (CDMFT) on a 2 $\times$ 2 plaquette. The CDMFT self-consistency equations \cite{Kotliar01,Maier05rmp} are
\beqa
\bm{G}(\iwn) &=&
\displaystyle
\sum_{\bm{\tilde{k}}}
\left(
(\iwn+\mu)\bm{1}-\bm{\Sigma}(\iwn)
-
\bm{t}(\bm{\tilde{k}})
\right)^{-1},
\\
\bm{\mathcal{G}}_0^{-1}(\iwn) &=& 
\bm{G}^{-1}(\iwn)-\bm{\Sigma}(\iwn).
\eeqa
For the $N_c$ $=$ $2 \times 2$ plaquette CDMFT, the hopping matrix 
$\bm{t}(\bm{\tilde{k}})$ is defined via its matrix elements 
$\bm{t}_{ij}(\bm{\tilde{k}})$ $=$ 
$N_c^{-1} \sum_{\bm{k}} e^{\text{i}(\bm{k}+\bm{\tilde{k}})\cdot(\bm{X}_i-\bm{X}_j)}$ 
$\epsilon_{\bm{k}+\bm{\tilde{k}}}$, where $\bm{X}_i$ and $\bm{X}_j$ are the 
position vectors of cluster sites $i$ and $j$, $\bm{\tilde{k}}$ is in the 
reduced Brillouin zone, and the cluster momenta take the values $\bm{k}$ $=$ 
$(0,0)$, $(\pi,0)$, $(0,\pi)$, and $(\pi,\pi)$. All quantities, i.e. $\bm{t}$, 
the coarse-grained cluster Green function $\bm{G}$, the Weiss field 
$\bm{\mathcal{G}}_0$, and the cluster self-energy $\bm{\Sigma}$ are 
$N_c \times N_c$ matrices, and $\bm{1}$ is the unit matrix. In the following we
consider only paramagnetic solutions and the spin index is therefore 
suppressed. 

The self-consistency cycle is closed by solving the impurity problem, i.e. by 
calculating a new cluster Green function matrix $\bm{G}_{ij}(\tau) = 
-\langle\mathcal{T}_\tau c_i^{}(\tau)c_j^{\dagger}(0)\rangle_{S_{\text{eff}}}$ for a
given self-energy and the corresponding Weiss field. Here $S_{\text{eff}}$ 
denotes the effective action of the auxiliary Anderson impurity model, which is
solved by numerically exact continuous-time quantum Monte Carlo (QMC) 
simulations based on the expansion of $S_{\text{eff}}$ in the impurity-bath 
hybridization.\cite{Werner06,Werner06b,Gull11review} In contrast to single-site DMFT,\cite{Metzner89a,Georges92,Jarrell92} CDMFT takes short-ranged nonlocal correlations within the cluster into account. These nonlocal correlations are particularly important for two-dimensional systems.\cite{Lichtenstein00,Stanescu04,Parcollet04,Maier05,Macridin06,Kyung06,Haule07,Ohashi08,Sakai09,Liebsch09,Sordi10}

We employ the following strategy for obtaining information on the critical 
behavior at the continuous Mott transition: First, we calculate the double 
occupancy $D$ $=$ $N_c^{-1}\sum_{i=1}^{N_c} \langle n_{i\uparrow} n_{i\downarrow} 
\rangle$ as a function of $U$ for a fixed ratio $t/T$ and search for 
hysteresis, i.e. whether there is a finite $U$ region in which both a metallic 
and a Mott-insulating solution of the self-consistent CDMFT equations exist 
depending on the initial guess for the self-energy. If hysteresis occurs, the 
temperature $T$ is increased, otherwise $T$ is decreased. This procedure is 
repeated until the boundary between hysteretic and non-hysteretic behavior is 
determined accurately. The critical inverse temperature for the continuous Mott
transition is denoted as $(t/T)_c$; the critical end point is determined by the
two parameters $(U/T)_c$ and $(t/T)_c$. Spin and charge criticality are 
subsequently measured and quantified by
\beqa
\sigma_{ch} &=& G_{\text{loc}}(1/(2T)),\\
\sigma_{sp} &=& \lim_{\omega \rightarrow 0} 
\frac{\text{Im}\;\chi_{\text{loc}}(\omega)}{\omega},
\eeqa
which are both functions of the reduced variables $t_{\text{red}}$ and 
$p_{\text{red}}$. 

The local Green function $G_{\text{loc}}(\tau)$ $=$ $N_c^{-1}\sum_{i=1}^{N_c} 
\langle c_i(\tau) c_i^{\dagger} \rangle$ measured at imaginary time $\tau$ $=$ 
$1/(2T)$ approximates $T A(\omega=0)$ and thus gives an estimate for the local 
density of states at the Fermi energy without necessitating an analytical 
continuation procedure for the imaginary-time data.\cite{Trivedi95,Gull08} Therefore 
$\sigma_{ch}$ serves as one possible measure for the criticality of the charge 
degree of freedom. The spin excitation spectrum is reflected in the local 
dynamical spin susceptibility $\chi_{\text{loc}}(\omega)$, which is calculated 
by a QMC measurement of the imaginary time correlation function 
$\chi_{\text{loc}}(\tau) = N_c^{-1}\sum_{i=1}^{N_c} \langle S_{i,z}(\tau) 
S_{i,z}(0)\rangle$ and the analytic continuation of its Matsubara transform to 
real frequencies. Here we use the maximum entropy method \cite{Jarrell96} for 
the bosonic kernel according to 
\beq
\chi_{\text{loc}}(\tau)=\int \frac{\text{d}\omega}{\pi} 
\frac{e^{-\tau\omega}}{1-e^{-\omega/T}} \text{Im}\;\chi_{\text{loc}}(\omega).
\eeq
\begin{figure}[htp]
\begin{center}
\includegraphics[width=0.95\hsize]{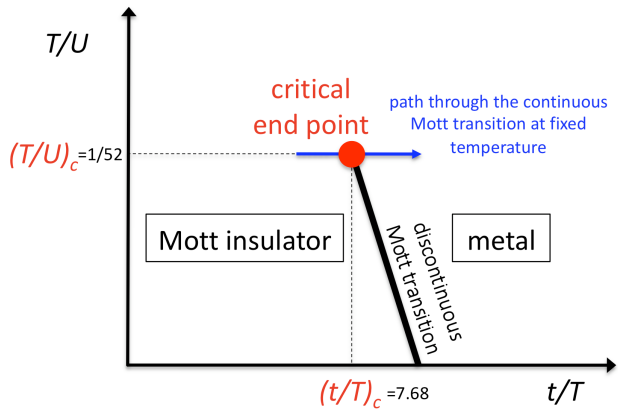}
\end{center}
\caption{
Schematic path through the continuous Mott transition at fixed temperature. We 
assume that the pressure-controlled transition in Ref.\ \onlinecite{Kagawa09} 
corresponds approximately to the bandwidth-controlled transition in the Hubbard
model. The hopping amplitude $t$ (bandwidth $W$ $\propto$ $t$) is varied (blue 
arrow) while the local Coulomb repulsion $U$, the relative anisotropic diagonal
hopping $t_{\text{diag}}/t$ and the temperature $T$ are kept fixed in our 
calculations. The indicated values of $(T/U)_c$ and $(t/T)_c$ are the critical 
values found for $t_{\text{diag}}/t$ = 0.44.}
\label{fig5_route}
\vspace{-4mm}
\end{figure}
In the Mott insulator ${\text{Im}\;\chi_{\text{loc}}(\omega)}/{\omega}$ is 
sharply peaked at $\omega$ $=$ 0. This behavior prohibits a reliable 
determination of $\sigma_{sp}$ in the insulator from our numerical data. We 
therefore restrict the analysis of spin criticality to the metallic side of the
transition. 

In order to model the bandwidth-controlled Mott transition by tuning the 
pressure, some further assumptions are necessary: We assume that varying $t/U$ 
amounts to varying pressure, and that $T/U$ is kept fixed at constant 
temperature. In essence this implies that the value of $U$ is fixed independent
of external conditions in the experiment. This is motivated by the fact that 
the Hubbard interaction is strictly local. Moreover we make the approximation 
that $t_{\text{diag}}/t$ = 0.44 remains fixed at the value taken from a fit to 
the band structure \cite{Kandpal09} even when pressure is applied. Fig.\ 
\ref{fig5_route} summarizes our strategy for modeling the continuous Mott 
transition across the critical end point.

\begin{figure}[htp]
\begin{center}
\vspace{2mm}
\includegraphics[width=\hsize]{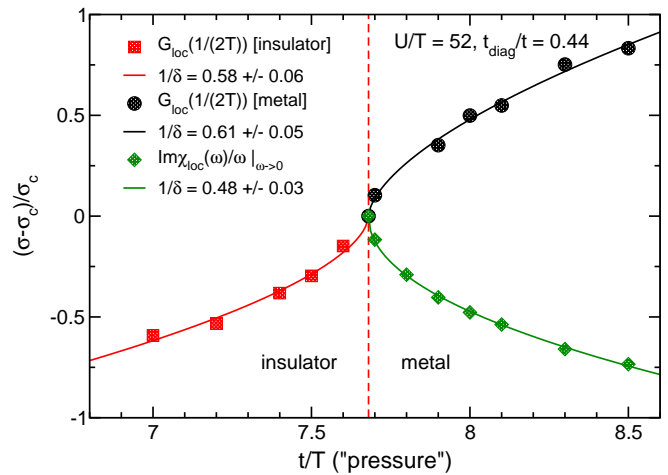}\\
\vspace{8mm}
\includegraphics[width=\hsize]{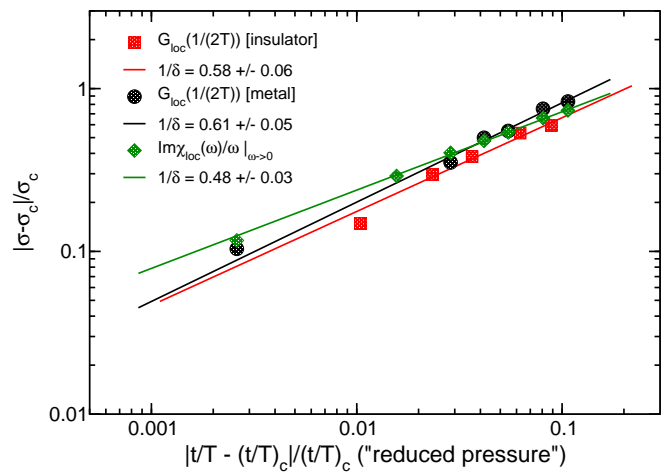}
\end{center}
\vspace{-4mm}
\caption[Charge and spin criticality]{
Critical behavior at the continuous bandwidth-controlled Mott transition 
(vertical dashed line, $(T/t)_c$ $=$ 7.68). Upper panel: Evolution of 
$(\sigma_{ch}-\sigma_{c,ch})/\sigma_{c,ch}$ and $(\sigma_{sp}-\sigma_{c,sp})/
\sigma_{c,sp}$ upon increasing $t/T$. The solid curves are fits to the critical
scaling behavior for the exponent $\delta$ according to Eq.\ 
(\ref{eq:exponents}). Lower panel: Double-logarithmic plot of the same data 
measured from the critical end point of the Mott transition. Solid lines show 
the scaling fits (the same fits as in the upper panel). Error bars are 
estimated from the linear regression fit to the data in the double-logarithmic 
scale.}
\label{fig5_scaling}
\vspace{-4mm}
\end{figure}
The basis for the discussion of the critical behavior are the data displayed in
Fig.\ \ref{fig5_scaling}. Both charge and spin degrees of freedom show critical
behavior with an infinite slope at the continuous transition, which is 
identified at $(t/T)_c$ $=$ 7.68, independently for charge and spin. The 
theoretical results resemble the experimental data for the criticality of the conductivity and the NMR spin-lattice relaxation rate $1/T_1$ in Ref.\ \onlinecite{Kagawa09} qualitatively. The increase of $G_{\text{loc}}(1/(2T))$ indicates that the low-energy spectral weight
increases upon passing from the insulator to the metal. This behavior reflects 
the closing and the filling of the charge gap with low-energy states at finite 
temperatures. Similarly, the measured conductivity will increase in the metal. 
In contrast, the spin susceptibility behaves oppositely; it is suppressed with 
increasing ``pressure'' on the metallic side of the continuous MIT and enhanced
in the insulator, as measured by $1/(T_1T)$.

This qualitative behavior of the spin susceptibility finds a natural 
interpretation in terms of the probabilities of relevant plaquette eigenstates in 
the ensemble.\cite{Haule07} The enhancement of spin fluctuations in the 
insulator is thereby traced to the predominant occupation of the plaquette with
a four-electron singlet state\cite{Gull08} with zero total momentum and zero total spin $S$ 
$=$ 0. The second-most probable states are the three triplet states with spin 
$S$ $=$ 1. Since the singlet state has a high occupation probability in the 
insulator, spin flip ($\Delta$ $S$ $=$ 1) excitations to the triplet states are
more likely and lead to the large susceptibility in the insulator. 

\begin{table}[htdp]
\caption{Summary of critical exponent $\delta$ for various models and experiments.}
\begin{center}
\begin{tabular}{|c|c|c|}
\hline
Model & $\delta$ & Ref.\\ 
 \hline
Ising ($D=\infty$) & 3 & [\onlinecite{Goldenfeld}] \\
Ising ($D=3$) & 4.8 & [\onlinecite{Goldenfeld}] \\
Ising ($D=2$) & 15 & [\onlinecite{McCoy73}] \\
Hubbard (single-site DMFT) & 3 & [\onlinecite{Kotliar00}] \\
Hubbard ($2 \times 2$ CDMFT) & & \\
$\sigma_{ch}$ (ins.) & 1.72 $\pm$ 0.17 & \\
$\sigma_{ch}$ (met.) & 1.64 $\pm$ 0.13 & \\
$\sigma_{sp}$ (met.) & 2.08 $\pm$ 0.10 & \\
\hline
Experiment & $\delta$ & \\
\hline
(V$_{1-x}$Cr$_x$)$_2$O$_3$ & $\approx$ 5 & [\onlinecite{Limelette03}] \\
$\kappa$-Cl & & \\
conductivity & $\approx$ 2Ê& [\onlinecite{Kagawa09}] \\
NMR $1/(T_1T)$ & $\approx$ 2 & [\onlinecite{Kagawa09}] \\
\hline
\end{tabular}
\end{center}
\label{table}
\end{table}%
The lower panel of Fig.\ \ref{fig5_scaling} shows the critical scaling behavior
in a double-logarithmic plot of $\sigma_{ch}$ and $\sigma_{sp}$ relative to 
their values at the critical point as a function of $|t/T-(t/T)_c|/(t/T)_c$. 
From linear fits to the double-logarithmic plot we extract the critical 
exponents according to Eq.\ (\ref{eq:exponents}). Error bars are estimated from
the linear regression fit to the data in the double-logarithmic scale. For the 
charge criticality the exponent $1/\delta$ $=$ 0.58 $\pm$ 0.06 is obtained on 
the insulating and $1/\delta$ $=$ 0.61 $\pm$ 0.05 on the metallic side of the 
transition. The critical behavior with an infinite slope of the density of 
states at the Fermi energy is apparent and resembles the measured dc conductivity.\cite{Kagawa05} The virtue of a direct quantitative comparison is, however, 
uncertain because the conductivity and the density of states at the Fermi 
energy may follow different scaling laws.\cite{Papanikolaou08} For the 
spin criticality, instead, a comparison is meaningful and the quantitative 
agreement with experiment is remarkable, with an exponent 0.48 $\pm$ 0.03 consistent within error bars with the experimentally determined $1/\delta$ $\approx$ 0.5.\cite{Kagawa09}

Table \ref{table} summarizes the values of the critical exponent $\delta$ $>$ 
$1$ for various theoretical models and for several experiments probing the 
continuous Mott transition. Note that $\delta$ $\rightarrow$ 1 corresponds to a
smooth transition with a vanishing discontinuity in the first derivative of 
$\sigma_{ch}$, while $\delta$ $\rightarrow$ $\infty$ describes a discontinuous 
transition with a discontinuity in $\sigma_{ch}$ itself. 

Our results for plaquette CDMFT are consistent with the experimental data for 
$\kappa$-Cl both for the charge and the spin degree of freedom. In particular, 
the critical exponent $\delta$ is much smaller than expected for the Ising 
universality class in two dimensions ($\delta$ $=$ 15) and also smaller than in
single-site DMFT ($\delta$ $=$ 3). The difference between single-site DMFT and 
four-site cluster DMFT is indeed striking. The latter includes nonlocal 
correlations and thereby allows for a coarse momentum-space differentiation. 
Hence, electrons in parts of the Brillouin zone become localized already in the
metallic phase. In contrast to single-site DMFT, where the Mott transition occurs 
simultaneously by a loss of quasiparticle integrity along the whole Fermi 
surface, the Fermi surface itself may disintegrate and ultimately vanish at the
Mott transition within cluster DMFT.\cite{Macridin06} Therefore, only a smaller fraction of 
electrons indeed exhibits critical behavior, which is why the transition is 
smoother (translating into a smaller value of $\delta$) than in single-site 
DMFT.

We acknowledge discussions with Fumitaka Kagawa and Kazushi Kanoda. This work 
is supported by the DFG through TRR 80. M.S. acknowledges support by the 
Studienstiftung des Deutschen Volkes. E.G. acknowledges support by 
NSF-DMR-1006282. Computer simulations were performed on HLRB II at LRZ Garching
using a code based on the ALPS libraries.\cite{alps3}
\bibliography{mybib}{}
\bibliographystyle{apsrev}
\end{document}